# Diffraction control in $\mathcal{PT}$-symmetric photonic lattices: from beam rectification to dynamic localization

Yaroslav V. Kartashov,[1,2] Victor A. Vysloukh,[3] Vladimir V. Konotop,[4] and Lluis Torner[1,5]

[1]*ICFO-Institut de Ciencies Fotoniques, The Barcelona Institute of Science and Technology, 08860 Castelldefels (Barcelona), Spain*
[2]*Institute of Spectroscopy, Russian Academy of Sciences, Troitsk, Moscow, 142190, Russia*
[3]*Departamento de Fisica y Matematicas, Universidad de las Americas Puebla, 72820 Puebla, Mexico*
[4]*Centro de Física Teórica e Computacional and Departamento de Física, Faculdade de Ciências, Universidade de Lisboa, Campo Grande 2, Edifício C8, Lisboa 1749-016, Portugal*
[5]*Universitat Politecnica de Catalunya, 08034, Barcelona, Spain*

**We address the propagation of light beams in longitudinally modulated $\mathcal{PT}$-symmetric lattices, built as arrays of couplers with periodically varying separation between their channels, and show a number of possibilities for efficient diffraction control available in such non-conservative structures. The dynamics of light in such lattices crucially depends on the ratio of the switching length for the straight segments of each coupler and the longitudinal lattice period. Depending on the longitudinal period, one can achieve either beam rectification, when the input light propagates at a fixed angle across the structure without diffractive broadening, or dynamic localization, when the initial intensity distribution is periodically restored after each longitudinal period. Importantly, the transition between these two different propagation regimes can be achieved by tuning *only* gain and losses acting in the system, provided that the $\mathcal{PT}$-symmetry remains unbroken. The impact of Kerr nonlinearity is also discussed.**

Periodic potentials can cause unidirectional motion of wavepackets along a direction that may be independent of the initial conditions. Two mechanisms leading to such a phenomenon have been discussed. One mechanism relies on a longitudinally (or temporal) modulated gradient imposed on transversally periodic lattices, which causes either dynamic localization or directional motion of the wavepacket, [1] depending on the modulation frequency. In Optics, such systems have been implemented as curved waveguide arrays [2-5], since the curvature of the array is equivalent to the presence of a gradient. Dynamically-varying lattices can be induced by interfering plane-waves [6,7]. Recently, directional beam motion, or *light rectification,* was observed in waveguide arrays with an out-of-phase varying curvature [8,9]. Such effect, termed *rectification* because the beam propagation direction becomes independent of the initial angle, relies on the dynamical band suppression due to the modulation of the effective coupling constants. A second mechanism, termed *ratchet,* that leads to directional motion is based on the concept of broken space-time symmetry [10] caused by specific lattice modulations. Ratchet schemes have been explored in the context of cavity solitons in coupled optical resonators [11]. Nonlinear dynamical settings offer especially rich opportunities for generating transverse motion [12].

New opportunities for diffraction control in lattices appeared with the advent of non-conservative guiding in structures obeying parity-time symmetry, which afford a stable balance between gain and losses [13]. Such systems support stationary guided modes that do not grow or decay upon propagation if the magnitude of the gain-losses does not exceed a critical level at which the $\mathcal{PT}$-symmetry is said to be broken [14], a phenomenon that was demonstrated experimentally in [15,16].

In this context, the spatial modulation of $\mathcal{PT}$-symmetric structures *along* the direction of light propagation opens an important new range of possibilities. In particular, besides dynamic localization that is possible in modulated lattices with unbroken $\mathcal{PT}$-symmetry [17], longitudinal variations of the system parameters may induce transitions between states with broken and unbroken symmetry [18]. Other examples of the phenomena afforded by the spatial modulations that have been uncovered are Kapitza stabilization in imaginary oscillating potential [19], resonant coupling of modes in multimode $\mathcal{PT}$-symmetric waveguides [20], realization of the pseudo-$\mathcal{PT}$-symmetry [21], stochastic parametric amplification in randomly modulated couplers [22], formation of self-sustained nonlinear modes [23] and their interaction with exceptional points (defects) [24], to name a few. However, the possibility of *controllable, unbounded transverse motion of wavepackets in a spatial $\mathcal{PT}$-symmetric lattice* has not been addressed so far. The study of such a phenomenon is the goal of this paper.

We uncover rich opportunities for simultaneous control of both, *diffraction broadening* and *transverse displacement* of wavepackets offered by dynamic $\mathcal{PT}$-symmetric lattices built as arrays of $\mathcal{PT}$-symmetric couplers with periodically varying separation between channels (Fig. 1). Depending on the ratio between the longitudinal period of the structure and the coupling length for straight segments of the individual couplers, one observes either *light rectification,* i.e. displacement of the beam in the transverse plane without broadening, or *dynamic localization,* when the input distribution is restored after each longitudinal period. Most importantly, we find that the transition between these two regimes can be achieved by varying only the depth of the dissipative part of the lattice, a quantity that allows flexible experimental control. We show the resonant character of the effect and illustrate higher-order localization or rectification resonances [25].

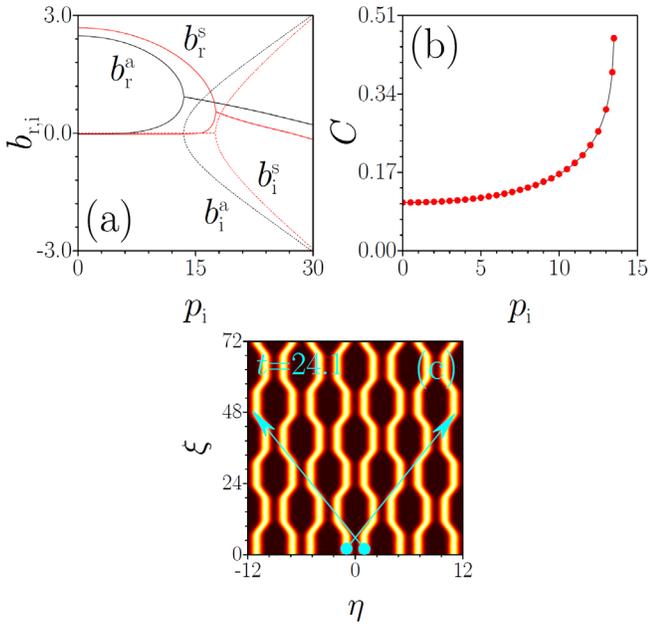

Figure 1. (Color online) (a) Real and imaginary parts of the propagation constants of modes supported by a two-channel straight $\mathcal{PT}$-symmetric structure versus $p_i$. (b) Coupling constant defined as $(b_r^s - b_r^a)/2$ versus $p_i$. The red dots show $C$ calculated from direct propagation. (c) Real part of the refractive index in a dynamic lattice with $t = 24.1$. The blue arrows indicate drift direction in one of the exact resonances. In all cases $\sigma = 0$. All quantities are plotted in dimensionless units.

We consider propagation of light in a $\mathcal{PT}$-symmetric lattice, whose refractive index as well as gain-losses profile are modulated both in the transverse, $\eta$, and in the longitudinal, $\xi$, directions. The evolution of the dimensionless field amplitude $q$ is governed by the nonlinear Schrödinger equation [2]:

$$i\frac{\partial q}{\partial \xi} = -\frac{1}{2}\frac{\partial^2 q}{\partial \eta^2} - \mathcal{R}(\eta,\xi)q - \sigma q |q|^2. \quad (1)$$

Here the transverse and longitudinal coordinates $\eta, \xi$ are scaled to the characteristic transverse scale $x_0$ and diffraction length $L_{\text{dif}} = k_0 x_0^2$, respectively, with $k_0 = 2\pi n/\lambda$ being the wave number and $n$ being the unperturbed background refractive index. The complex refractive index distribution

$$\mathcal{R}(\eta,\xi) = \sum_{k=-\infty}^{\infty} \{\mathcal{R}_0[\eta + \eta_c(\xi) + 3kd] + \mathcal{R}_0[\eta - \eta_c(\xi) + 3kd]\} \quad (2)$$

involves two shifted sublattices with opposite directions of waveguide bending in $\xi$, where $\mathcal{R}_0 = \mathcal{R}_r + i\mathcal{R}_i$ is the refractive index profile of a single waveguide in the array, whose real and imaginary parts are given by $\mathcal{R}_r = p_r \exp(-\eta^2/a^2)$ and $\mathcal{R}_i = p_i \eta \exp(-\eta^2/a^2)$, respectively. Here the parameters $p_{r,i} \sim \delta n_{r,i} k_0^2 x_0^2$ define the modulation depths of the real and imaginary parts of the refractive index; $3d$ is the transverse period of the structure; and $a$ is the waveguide width. Notice that within each waveguide the refractive index is symmetric, while gain-losses are antisymmetric, hence each sublattice and the entire structure are $\mathcal{PT}$-symmetric: $\mathcal{R}_0(\eta,\xi) = \mathcal{R}_0^*(-\eta,\xi)$. The positions of guides vary with distance $\xi$ in accordance with the periodic differentiable function $\eta_c(\xi) = \eta_c(\xi + 4l)$, where $t = 4l$ is the longitudinal period. Such structure involves two straight segments with $\eta_c = d/2$ for $\xi \in [0, l]$ and $\eta_c = d$ for $\xi \in [2l, 3l]$ that are connected by two smooth transitions $\eta_c = (d/2)[1 + \cos^2(\pi\xi/2l)]$ for $\xi \in [l, 2l]$ and $\eta_c = (d/2)[1 + \sin^2(\pi\xi/2l)]$ for $\xi \in [3l, 4l]$. This choice of lattice shape is not unique, of course, and different segments may have different lengths. A representative distribution of the real part of the refractive index is depicted in Fig. 1(c). Such a structure can be viewed as an array of couplers with a periodically varying separation between channels between $d$ and $2d$. We set the values of the parameters to minimize the radiative losses caused by the channel bending. Equation (1) accounts also for focusing ($\sigma = +1$) or defocusing ($\sigma = -1$) Kerr nonlinearity.

The gain/loss landscapes for modulated $\mathcal{PT}$-symmetric lattices can be realized using inhomogeneous doping of a host material with active centers (active ions) and a proper illumination of the sample, which can also be made spatially inhomogeneous. Such doping is a well-developed technology in photorefractive crystals, where optical induction is routinely used to create dynamical refractive index landscapes [26]. Electric pumping using properly shaped surface electrodes also makes it possible to realize inhomogeneous gain. The strength of the imaginary part of the lattice can be tuned by the intensity of the pump beam, as shown in [16], thereby allowing full control of the effective coupling in structures with a fixed geometry, as shown below. Modern fabrication technologies also allow to achieve different level of losses in complex guiding structures: one example of such technique not relying on doping was introduced recently in laser-written waveguide arrays [27]. The possibility to tune effective coupling by varying gain/loss level is a clear advantage in comparison with conservative settings, where realization of new propagation regimes requires manufacturing a new lattice. This makes it possible to demonstrate the transition from rectification to dynamic localization *without changing the geometry of the structure*. Furthermore the structure depicted in Fig. 1(c) allows observation of higher-order rectification and dynamic localization resonances, which have never been demonstrated so far.

Light propagation in a $\mathcal{PT}$-symmetric dynamic lattice strongly depends on the ratio between the coupling length defined for straight, closely-spaced lattice segments and the length $l$ of these segments (or total period $t = 4l$). This can be seen from the tight-binding approximation of Eq. (1) that holds in the linear (i.e. low-power) limit with $\sigma = 0$. Since the array is formed by two sublattices with identical waveguides, the field distribution can be written as $q(\eta,\xi) = e^{ib\xi} \sum_k [a_k^{(1)}(\xi)\phi(\eta + \eta_c + 3kd) + a_k^{(2)}(\xi)\phi(\eta - \eta_c + 3kd)]$ in terms of the eigenmodes $\phi(\eta,\xi)$ of the individual waveguides $\mathcal{R}_0(\eta,\xi)$ forming the sublattices [i.e., $\phi$ satisfies $(1/2)d^2\phi/d\eta^2 + \mathcal{R}_0\phi = b\phi$, where $b$ is the propagation constant]. We assume that the $\mathcal{PT}$-symmetry is unbroken ($b$ is real), substitute the above expression for the field into Eq. (1) and project (1) over the modes $\phi^*(\eta \pm \eta_c + 3kd)$, representing modes of the $\mathcal{R}_0^*(\eta,\xi)$ potential. Following the standard procedure, where only nearest-neighbor coupling is taken into account, one obtains the discrete equations for mode weights $a_k^{(1)}, a_k^{(2)}$ in each sublattice:

$$ida_k^{(1)}/d\xi = -\omega(\xi)a_k^{(1)} - C_1(\xi)a_k^{(2)} - C_2^*(\xi)a_{k-1}^{(2)},$$
$$ida_k^{(2)}/d\xi = -\omega^*(\xi)a_k^{(2)} - C_1^*(\xi)a_k^{(1)} - C_2(\xi)a_{k+1}^{(1)},\quad(3)$$

with $\xi$-dependent coefficients

$$\omega = \frac{1}{Q}\int_{-\infty}^{\infty}\phi^2(\eta+2\eta_c)[\mathcal{R}_0(\eta)+\mathcal{R}_0(\eta+3d)]d\eta,$$
$$C_{1,2} = \frac{1}{Q}\int_{-\infty}^{\infty}\phi(\eta-2\eta_{1,2})[\phi(\eta)\mathcal{R}_0(\eta)-i\eta_c'd\phi(\eta)/d\eta]d\eta,\quad(4)$$
$$Q = \int_{-\infty}^{\infty}\phi^2(\eta)d\eta,$$

where $\eta_1 = \eta_c$, $\eta_2 = \eta_c - 3d/2$. The quantity $Q$ in (4) is real because of the $\mathcal{PT}$-symmetry of the modes $\phi(\eta) = \phi^*(-\eta)$ and for the same reason, the contribution to $C_{1,2}$ from the term $\sim \eta_c' = d\eta_c/d\xi$ is real too. The effective gain and losses stemming from the imaginary part of $\omega(\xi)$ only weakly affect the local dynamics and disappear after averaging over the propagation distance $\xi$, since $\omega(\xi) = \omega^*(\xi+2l)$. Moreover, the term $\sim d\eta_c/d\xi$ contributes to the coupling constants $C_{1,2}$ only on the lattice segments with nonzero curvature, where waveguides are mostly well separated and $|C_{1,2}|$ is relatively small (hence power transfer is negligible). Thus, intense power transfer between guides occurs mostly on straight lattice segments, where $\eta_c$ is constant and only one of the coefficients $C_{1,2}$ has a non-negligible value $C = \max C_{1,2}$ (due to considerable difference of waveguide separation in a given coupler and distance to waveguides in the neighboring coupler), which is given by the expression $C = Q^{-1}\int_{-\infty}^{\infty}\phi(\eta)\mathcal{R}_0(\eta)\phi(\eta-d)d\eta$.

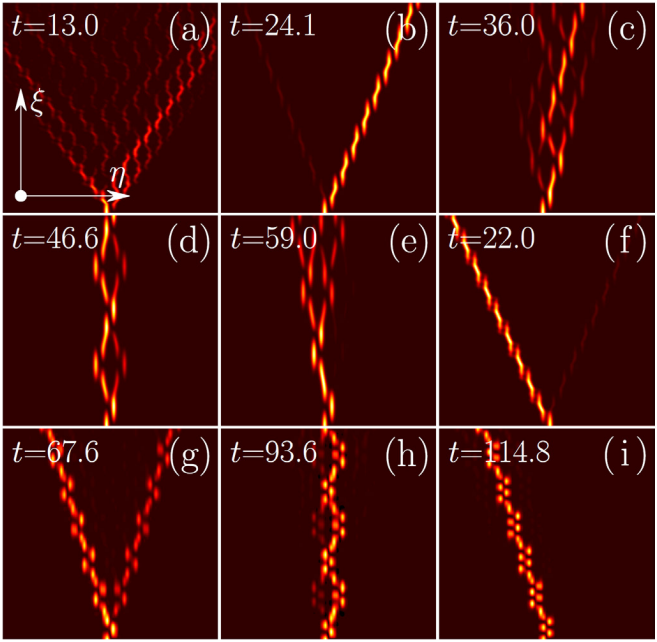

Figure 2. (Color online) Evolution dynamics in a linear $\mathcal{PT}$-symmetric dynamical lattice, for different longitudinal periods at $p_i = 10$, $\sigma = 0$. In (a)-(e) and (g)-(i) the left channel of central pair is excited at $\xi = 0$, while in (f) the right channel is excited. The propagation distance in (a)-(f) is $\xi = 100$, while in (g)-(i) it is equal to $4t$. The intensity distributions are shown within the $\eta \in [-30, +30]$ window.

This coupling constant strongly depends on the depth of the imaginary part of the refractive index. Fig. 1(b) shows the $C(p_i)$ dependence calculated for the simplest possible structure, namely for two straight waveguides separated by the distance $d$. A pair of waveguides supports symmetric and antisymmetric linear guided modes when $p_i = 0$. The evolution of their complex propagation constants $b^{a,s} = b_r^{a,s} + ib_i^{a,s}$ versus $p_i$ is shown in Fig. 1(a). Further we set $p_r = 5$, $d = 2$, $a = 1/2$. For such parameters, $\mathcal{PT}$-symmetry breaking occurs at $p_i \approx 13.5$, when the eigenvalue of the antisymmetric mode becomes complex (after collision with the eigenvalue of a new mode whose propagation constant shifts from the continuum to the discrete spectrum with increasing $p_i$), even though the symmetric mode possesses a real propagation constant up to $p_i \approx 17.6$. Note that at $p_i \approx 17.6$, the same scenario is encountered for the symmetric mode. The effect closely resembles the collision of the propagation constants of pairs of modes in multimode $\mathcal{PT}$-symmetric waveguides described in [20]. The mechanism that creates this effect is that internal currents in a non-conservative system may act towards the equilibration of the field modulus distributions of modes, which in conservative potentials have totally different symmetries. The coupling constant can be defined as half of the difference $C \approx (b_r^s - b_r^a)/2$ of the propagation constants of the highest symmetric and antisymmetric modes. It grows rapidly when $p_i \to 13.5$, but remains finite at the symmetry breaking point, where the tangential line to $C(p_i)$ dependence becomes vertical. This dependence is confirmed by the direct solution of Eq. (1) indicated by red dots in Fig. 1(b).

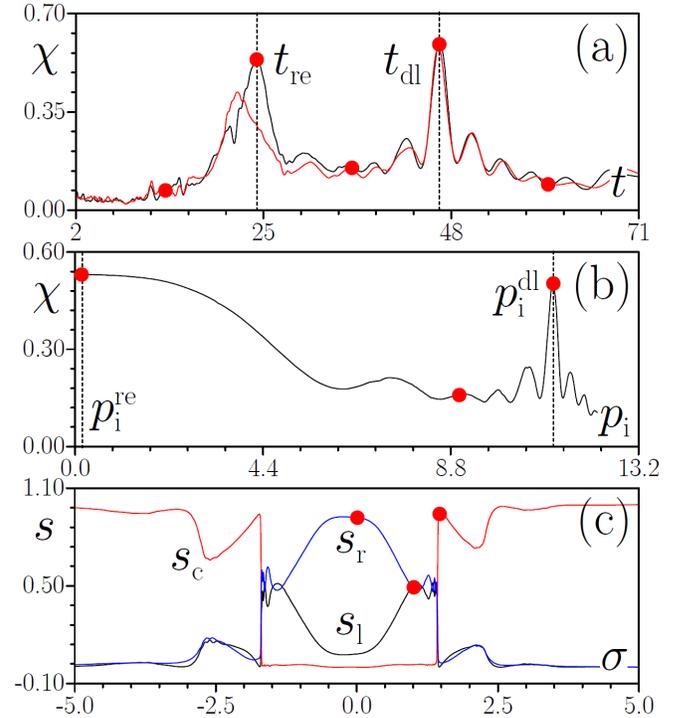

Figure 3. (Color online) (a) Output form-factor as a function of the longitudinal period of the structure at $p_i = 10$ for excitation of the left (black curve) and right (red curve) waveguides in the input pair at

$\sigma = 0$. (b) Output form-factor versus $p_i$ at $t = 39$, $\sigma = 0$ for excitation of the left waveguide. In (a), (b) dashed lines indicate parameters corresponding to rectification or dynamic localization. (c) Output power concentrated if left $s_l$ and right $s_r$ parts of the array and in central waveguide pair $s_c$ versus $\sigma$ at $p_i = 5$, $t = 34.5$. Red dots correspond to dynamics from Figs. 2 and 5. All quantities are plotted in dimensionless units.

Next we address the full dynamic lattice [Fig. 1(c)]. For all results shown below we use as input for Eq. (1) the superposition $\phi^s \pm \phi^a$ of symmetric and antisymmetric modes of the central waveguide pair, taken with equal weights. This reduces radiative losses at the initial stage of propagation and allows exciting either the left or the right waveguide in the central pair, depending on the sign in the superposition.

As discussed above, the dynamics of propagation even in a linear medium at $\sigma = 0$ strongly depends on the ratio of the coupling length $\pi/C$ for the straight segments and on the length $l$ of those segments (period $t = 4l$). If $l$ is small, one observes strong diffraction [Fig. 2(a)]. If $l$ is selected such that at the straight segments the beam completely switches to the neighboring waveguide, then light couples to the adjacent pair, where the process is repeated, leading to directional transport (*rectification*) across the lattice, almost without diffraction [Fig. 2(b)]. The transport direction depends on which waveguide is initially excited in the central pair and the corresponding optimal periods are slightly different due to intrinsic anisotropy introduced by the dissipative part of the $\mathcal{PT}$-symmetric lattice [Fig. 2(f)]. While for slightly larger periods the rectification effect is lost [Fig. 2(c)], when $l$ is approximately equal to two coupling lengths, one observes *dynamic localization* upon which the initial intensity distribution is restored after each period $t$, although light couples to the neighboring waveguide pairs [Fig. 2(d)]. A further increase of the longitudinal period reveals the rich structure of high-order rectification or localization resonances [Figs. 2(g)-2(i)], each of them being connected with an *odd* or *even* number of switching events on straight segments. Thus, the lattice addressed here makes it possible to observe two qualitatively different types of propagation dynamics.

Rectification can also be observed in lattices with individual guides of the form $\mathcal{R}_0 = (p_r \pm ip_i)\exp(-\eta^2/a^2)$ although in such a system the $\mathcal{PT}$-symmetry breaking threshold is substantially lower ($p_i = 0.16$ at $p_r = 5$), and variation of the coupling constant is not so pronounced. Therefore, the possibilities to tune drift angle are reduced.

To illustrate the resonant character of the phenomenon described above, it is convenient to use the integral form-factor $\chi = U^{-2} \int_{-\infty}^{\infty} |q|^4 d\eta$, where $U = \int_{-\infty}^{\infty} |q|^2 d\eta$ is the total power (a larger $\chi$ implies better localization of the output pattern). We also calculated power fractions contained in a central waveguide pair $s_c = \int_{-3d/2}^{3d/2} |q|^2 d\eta$, as well as in the left $s_l = \int_{-\infty}^{-3d/2} |q|^2 d\eta$ and right $s_r = \int_{3d/2}^{\infty} |q|^2 d\eta$ half-spaces, outside the central pair. The resonant character of the effect is clearly visible in Fig. 3(a), which shows the dependence of the output form-factor at $\xi = 5t$ on the longitudinal period of the structure for excitations of the left and right waveguides in the central pair. The first resonance peak at $t = t_{re}$ corresponds to rectification [see Fig. 2(b) for the corresponding dynamics], while the second one at $t = t_{dl}$ corresponds to dynamic localization [Fig. 2(d)]. Here we show only two resonances, but there are many of them and they alternate.

It is important to stress the obvious difference in the positions of the rectification resonances for excitations of left and right guides in the central pair (that disappears at $p_i \to 0$). In contrast, dynamic localization resonance is not sensitive to the position of the initial excitation. The dependencies of resonant periods for the first two resonances on $p_i$ are shown in Fig. 4(a) (in the case of rectification resonance we provide both resonant periods corresponding to $l \to r$ and $r \to l$ transport directions). All resonant periods monotonically decrease with $p_i$. In contrast, the rectification angle defined as $\alpha = d\eta_c / d\xi$, where $\eta_c = U^{-1} \int_{-\infty}^{\infty} \eta |q|^2 d\eta$ is the position of the integral center of the pattern, rapidly grows with $p_i$ which agrees with the fact that the growing imaginary lattice part results in tunneling enhancement [Fig. 1(b)]. The difference between $\alpha_{l \to r}$ and $\alpha_{r \to l}$ angles becomes most pronounced close to the symmetry breaking point indicated by the dashed line.

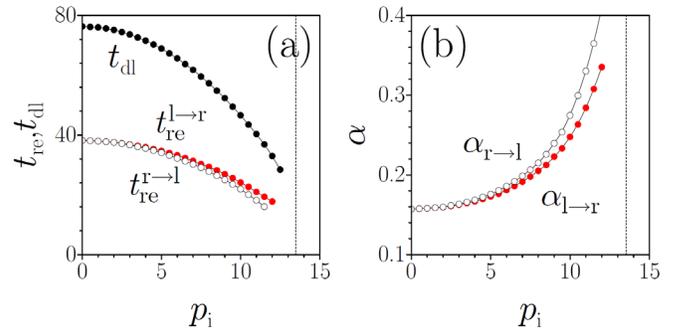

Figure 4. (Color online) (a) Optimal longitudinal periods of the structure corresponding to rectification ($t_{re}$) and dynamic localization ($t_{dl}$) versus depth of the imaginary part of the lattice. (b) Velocity of transverse displacement arising upon rectification versus $p_i$. Dashed lines indicate symmetry breaking point for a pair of isolated straight waveguides. In all cases $\sigma = 0$. All quantities are plotted in dimensionless units.

The key difference between conservative and $\mathcal{PT}$-symmetric lattices is linked with the possibility to control the light evolution by varying *exclusively* the strength of the gain-losses parameter $p_i$ in the $\mathcal{PT}$-symmetric structures. Figure 3(b) illustrates the variation of the output form-factor at $\xi = 5t$ upon increase of $p_i$ at a fixed longitudinal period $t$, while Figs. 5(a)-5(c) show the corresponding evolution dynamics. One can see that increasing $p_i$ leads to transition from *rectification* (broad resonance at $p_i \approx 0.4$) to *dynamic localization* (narrow resonance at $p_i \approx 11.2$) in the structure with fixed real refractive index landscape. Thus, variation of the strength of gain-losses not only has a quantitative impact (modification of propagation angles), but rather qualitatively changes the character of light propagation.

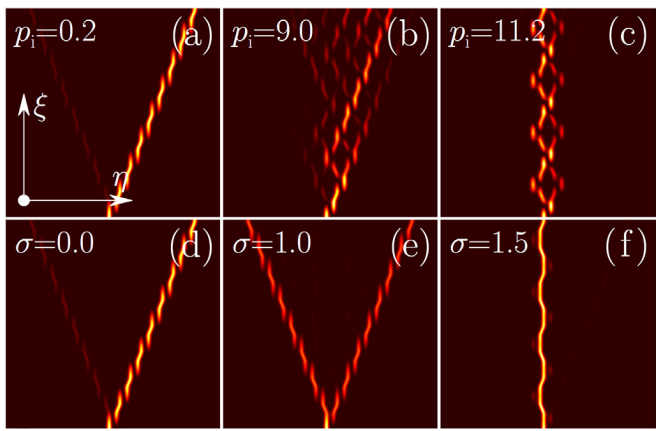

Figure 5. (Color online) (a)-(c) Modification of propagation dynamics in the dynamical lattice with $t=39$, $\sigma=0$ upon increase of $p_i$. (d)-(f) Modification of propagation dynamics in the lattice with $p_i=5$, $t=34.5$ upon increase of $\sigma$. In all cases the propagation distance is $4t$. Dynamics is shown within $\eta \in [-30,+30]$ window. Selected values of $p_i$ in (a)-(c) and $\sigma$ in (d)-(f) correspond to red dots in Fig. 3.

Finally, the discussed phenomena can be strongly affected by the presence of Kerr self-focusing or self-defocusing nonlinearity in the lattice. This is illustrated in Fig. 3(c) where we show the output power fractions $s_l, s_c, s_r$ concentrated in the left and right half-spaces as well as in the central waveguide pair as functions of the nonlinearity coefficient $\sigma$. The period $t$ was selected in such a way that when $\sigma=0$ one obtains light rectification for the excitation of the left guide, so that most of the power at $\xi=5t$ is concentrated in the right half-space [see Fig. 5(d)]. Increasing the nonlinearity (whether focusing or defocusing) results in the *equilibration* of intensities of the beams propagating in the left and right directions [Fig. 5(e)]. The effect is similar to the equilibration of output powers concentrated in the arms of a straight conservative coupler that occurs when the input power approaches the critical value for switching, which in turn decreases with a decrease of the coupling constant [28]. As expected on physical grounds, at sufficiently high nonlinearities (i.e., when the input power exceeds the critical value for switching) an abrupt transition to localization in the excited waveguide occurs [Fig. 5(f)].

In summary, we have addressed light rectification and dynamic localization phenomena in $\mathcal{PT}$-symmetric modulated optical lattices. We have shown that tuning of the light transport regimes occurs by adjusting *only* the strength of the gain-losses control parameter, thus without changing the lattice geometry as it would be required in conservative systems. The variety of propagation regimes encountered in the studied system emphasizes the important fundamental and technological opportunities for light diffraction control offered by non-conservative potentials.

**Acknowledgment**. This work was partially supported by the Severo Ochoa Excellence program SEV-2015-0522. VVK was supported by the FCT (Portugal) under the grants PTDC/FIS-OPT/1918/2012 and UID/FIS/00618/2013.